\documentclass{WileyMSP-template}

\usepackage{amsmath}
\usepackage{textcomp, gensymb}
\usepackage{soul}
\usepackage{hyperref}
\usepackage{url}
\usepackage{physics}
\usepackage[version=4]{mhchem} 
\usepackage{mathtools}
\usepackage{graphicx}% Include figure files
\usepackage{dcolumn}% Align table columns on decimal point
\usepackage{float}
\usepackage{bm}% bold math
\usepackage{subcaption} % it allows to have sub figs within a fig

\usepackage{xcolor}

\begin{document}

%\pagestyle{fancy}
%\rhead{\includegraphics[width=2.5cm]{vch-logo.png}}

\title{Studies of parity violation in atoms}

\maketitle

% Author: Please give full first and last names for authors and include * after the name of all corresponding authors

\author{Stefanos Nanos}
\author{Iraklis Papigkiotis}
\author{Dionysios Antypas*}

% Affiliations: Please provide adacemic titles (Prof. or Dr.) for all authors where applicable, and include an institutional email address for all corresponding authors
\begin{affiliations}
Dr. S. Nanos, I. Papigkiotis, Prof. D. Antypas\\
Department of Physics, University of Crete, GR-70013 Heraklion, Greece\\
Email Address: dantypas@physics.uoc.gr

\end{affiliations}

% Keywords: Please provide a minimum of three and a maximum of seven keywords, separated by commas


% Abstract should be written in the present tense and impersonal style (i.e., avoid we), and be at most 200 words long
\begin{abstract}

Studies of the effects of the weak interaction in atomic systems provide tests of the Standard Model of particle physics, and explore physics scenarios beyond  the Standard Model. In addition, these studies can offer valuable insights into low-energy nuclear physics. We provide an overview of the field of atomic parity violation, and discuss implications to nuclear and particle physics, and ongoing experimental efforts. Furthermore, we present our plans for precision measurements of the signatures of the weak interaction in atomic ytterbium.  
\end{abstract}

\section{Introduction}

 Investigating fundamental physics involves a range of approaches, each contributing uniquely to our understanding of the laws of physics. High-energy colliders \cite{Shi12, Gra21} have been at the forefront of particle physics, providing insights into the subatomic world. A recent milestone is the confirmation of the Higgs boson \cite{ATL12, CMS12}, which completed the Standard Model (SM) of particle physics and improved our understanding of how particles acquire mass \cite{Ber75}. 
 In addition, astrophysical probes provide  evidence that hint at physics beyond the SM. For instance, an abundance of direct observational evidence \cite{ArbeyPPNP2021} points to the presence of dark matter in the universe.
\paragraph{}
 Another approach to investigating fundamental physics is within the so-called low-energy, precision frontier, where high measurement precision is employed, typically in small-scale apparatus, to probe tiny deviations of observables from the SM predictions \cite{saf18}. Experiments employing atoms and molecules are particularly well-suited for such investigations due to the high precision achievable in their experimental manipulation and measurement. Such experiments may provide information complementary to that of high-energy apparatus.   
Related activities include tests of violations of fundamental symmetries in the SM. One example is the study of time-reversal symmetry violation via searches for permanent electric dipole moments (EDMs) of elementary particles\,\cite{ACM18, Rou22}, the observation of which would aid to the understanding of the matter-antimatter asymmetry in the universe\,\cite{dem03}. 
%\paragraph{}
Another example is the violation of symmetry with respect to spatial inversion (parity). This parity violation manifests itself in the weak interaction and, in atomic systems, arises either due to the neutral $Z^0$ boson exchange between electrons and nucleons or intranuclear weak interactions. The former interaction is nuclear-spin-independent (NSI) while the latter are nuclear-spin-dependent (NSD). Atomic parity violation (APV) induces observables on atomic states, measurements of which can provide checks of the electroweak SM theory and information about low-energy nuclear physics.
\paragraph{}
Shortly after Wu's discovery of parity violation in beta decay \cite{WuPR1957}, Zel'dovich pointed out that the electron-nucleus weak interaction is expected to induce a parity-odd (P-odd) atomic observable \cite{ZelDovich1959PARITYEFFECTS}. However, he considered hydrogen (atomic number $Z=1$), in which an APV observable would be too small for detection. Bouchiat and Bouchiat showed that the APV effect grows slightly faster than $Z^3$ \cite{Bouchiat1974WeakPhysics, BouchiatJPF1974, Bouchiat1975ParityII}, thus motivating experiments in heavy atoms. The first APV measurements in Bi \cite{Barkov1978ObservationTransitions} provided an early confirmation of the electroweak theory, years before the discovery of the weak-force mediators ($W^{\pm}$ and $Z^0$ bosons). Precise experiments (with accuracy greater than $5\%$ in the APV effect) were carried out in Bi \cite{MacPherson1991PreciseBismuth}, Pb \cite{Meekhof1993High-precisionLead, Phipp1996ALead}, Tl \cite{Vetter1995PreciseThallium}, Cs \cite{Wood1997MeasurementCesium, Guena2005MeasurementDetection} and Yb \cite{Ant19a}. The most precise measurement, done in Cs, reached an accuracy of $\approx 0.4\%$ and combined with precision atomic calculations \cite{dzu12}, it provided a stringent test of the SM at low energy, via the determination of the so-called weak charge $Q_W$ of the Cs nucleus, an electron-nucleus weak-coupling constant. Moreover, the Cs experiment has been the only one to date to provide an unambiguous determination of the so-called anapole moment, a P-odd electromagnetic (EM) moment arising due to intranuclear weak interactions \cite{Haxton2001AtomicMoments}.
\paragraph{}
The most recent APV measurements were done in Yb in a chain of isotopes \cite{Ant19a}, demonstrating directly the isotopic variation of the electron-nucleus weak interaction.  The electronic structure of Yb is not simple enough to allow for precision atomic calculations, so that (unlike in the Cs case) an accurate weak-charge extraction from measurements in a single isotope is currently unfeasible. However, one may take the ratio of APV in two isotopes, circumventing the need for  theory. This isotopic ratio method \cite{Dzuba1986EnhancementAtoms} can be used to probe nuclear neutron distributions \cite{Fortson1990Nuclear-structureNonconservation}, or physics beyond the SM \cite{Bro09}. Moreover, the availability of isotopes of Yb with nuclear spin allows for measurements of NSD  effects and the Yb nucleus anapole moment. 

 \section{Manifestation of Atomic Parity Violation}

\begin{figure}
\centering
\includegraphics[width=1\linewidth]{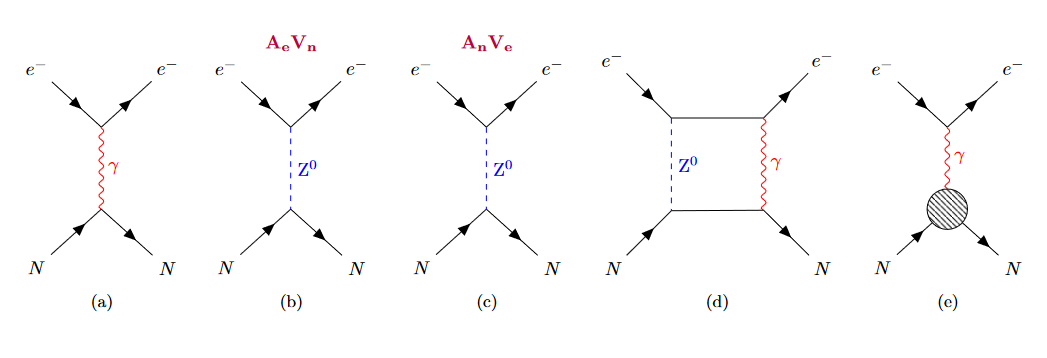}
\caption{Feynman diagrams (created with TikZ package \cite{Ell17}) mainly contributing to APV observables \cite{Roberts2015ParitySystems,  Haxton2001AtomicMoments}}
\label{fig:feynman}
\end{figure}

 The primary source of APV is due to the exchange of neutral $Z^0$ bosons between the electron and nucleus. Unlike the photon-mediated electromagnetic force, the weak force is a contact interaction on the atomic scale (interaction length $\hbar/M_Z c\approx 2\cdot 10^{-3}$\,fm), owing to the large $Z^0$ mass $M_{Z}\approx 91.2$\,GeV. Thus, only electrons in $s$-orbitals, which penetrate into the nucleus experience the weak force. Feynman diagrams for the electromagnetic and weak ($Z^0$-exchange) interactions are shown, respectively, in Fig.~\ref{fig:feynman}a and Fig.~\ref{fig:feynman}b. Perturbation from the weak interaction leads to a mixing between opposite parity $s$ and $p$ orbitals, and respective atomic states, therefore,  acquire a small, opposite-parity admixture.  This gives rise to a small, weak-force-induced electric-dipole (E1) transition amplitude $A_w$ in an otherwise `forbidden` atomic transition. To probe APV,  one employs a  transition with a moderate, EM-conserving transition amplitude $A_{\rm em}$ that is greater than $A_w$ [e.g. a magnetic-dipole (M1), or an E1 transition amplitude induced with a static electric field]. One provides  experimental conditions that induce interference between the two amplitudes \cite{Bouchiat1986OpticalInteractions}, resulting in a transition rate given by: 
 
 \begin{equation}
    \label{eq:transition_rate}
    |A_{em}+A_{w}|^2 = A_{em}^2   \pm 2 A_{em} A_{w} \;, 
\end{equation}
\\
where $A_{\rm em}$ and $A_{\rm w}$ are taken to be real, and the small term $ A_w^2$ has been neglected. The change in the transition rate due to this interference (represented by the last term in Eq.~\ref{eq:transition_rate}) shows the degree of parity violation between the two `mirror' versions of an experiment, and provides a measurable observable, determined  by normalizing the transition rate difference to the mean rate. This asymmetry is given by:

 \begin{equation}
    \label{eq:rl_asymmetry}
    \mathcal{A} = \frac{2A_{\mathrm{w}}}{A_{\mathrm{em}}}\;,
\end{equation}. 

and is enhanced, not only when $A_{\mathrm{w}}$ is large, but also in highly forbidden transitions (i.e. for small $A_{\rm em}$). In APV experiments it may be of order $10^{-7}-10^{-4}$.
\paragraph{}
To date, APV has been probed  in i) Stark-interference, and ii) optical rotation experiments. In the first method, an E1- and M1- forbidden transition is employed. The small-E1 amplitude arising due to APV-induced mixing interferes with an E1 amplitude introduced with a static electric field. Field reversals are made to switch the handedness of the experiment and create a modulation in the excitation rate. Stark-interference has been used in experiments done in Cs, Tl, Yb \cite{Wood1997MeasurementCesium,
LintzEL2007,
Conti1979PreliminaryThallium,
Tsigutkin2009ObservationYtterbium,
Ant19a}.
In the second method, the APV-induced E1 amplitude in a E1-forbidden but M1-allowed transition, interferes with the M1-amplitude. This interference induces circular dichroism in the atomic medium, and thus, optical rotation of the polarization plane of linearly polarized light exciting the atomic medium. This method was employed in measurements in Bi, Pb and Tl \cite{MacPherson1991PreciseBismuth, Meekhof1993High-precisionLead,
 WarringtonEL1993,
Edwards1995PreciseThallium,
Vetter1995PreciseThallium,
 Phipp1996ALead} .

\subsection{Nuclear-spin independent APV}

\begin{figure}{}
    \centering
    \includegraphics[width=0.7\linewidth]{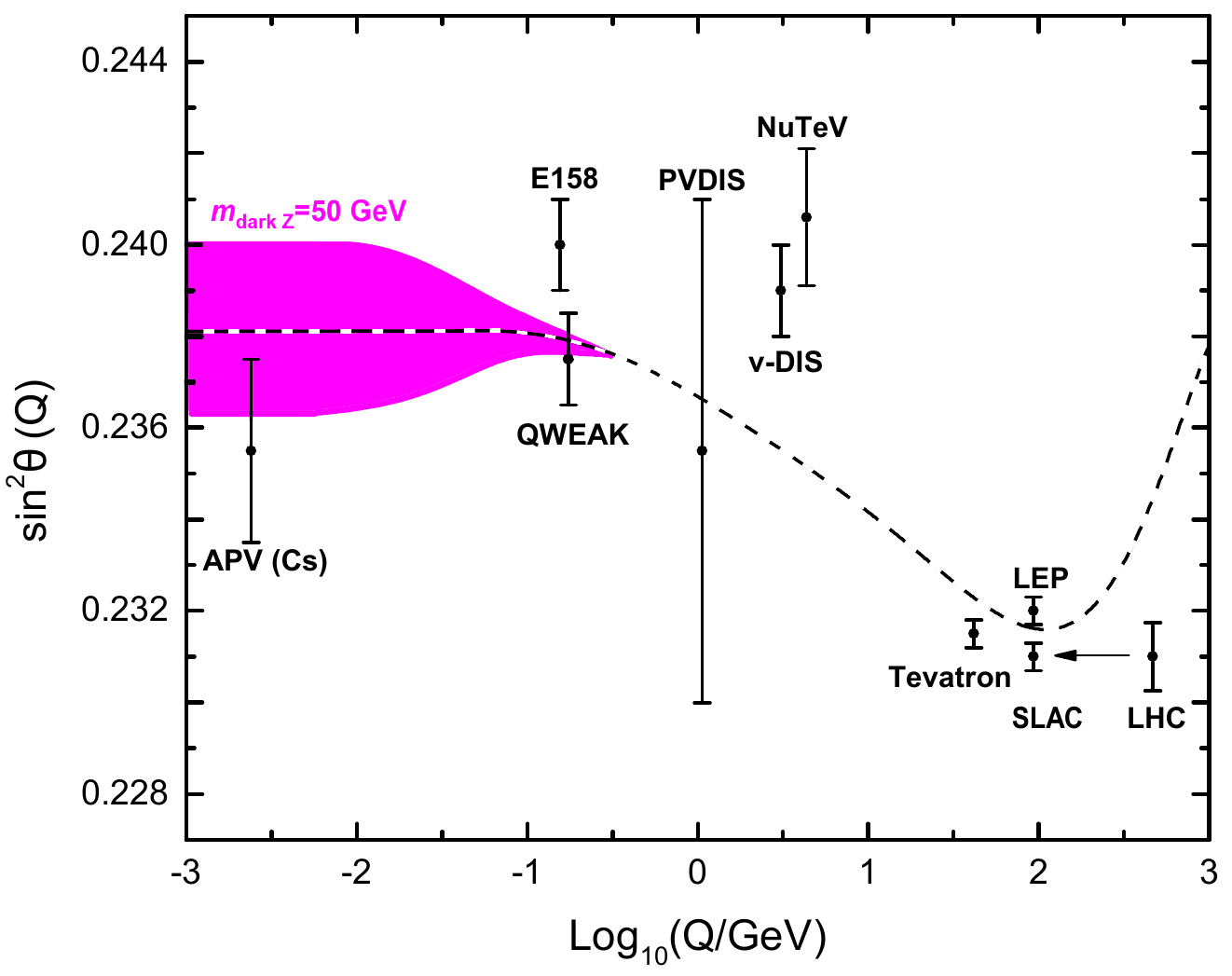}
    \caption{ Experimental determinations of $\sin^2{\theta_W}$ and SM prediction (dashed line) \cite{Accardi2016,dav15, AndroicNature2018-QWEAK}. The pink region shows the modification of $\sin^2{\theta_W}$ due to a dark $Z$ boson with mass $m_{dark\ Z}=50\  \mathrm{GeV}$ \cite{dav15,dav14}.}
    \label{fig:weinberg}
 \end{figure}

The NSI weak interaction between an electron and an infinitely heavy nucleon  is given by \cite{KhriplovichPNCbook}:%{\SN{:}}

\begin{equation}
\rm{H}=\frac{G_F}{\sqrt{2}} \mathit{C}_{1i} \gamma_5\delta(\mathbf{r}),
    \label{eq:Hamiltonian1}
\end{equation}

where $\mathrm{G_F}$ is the weak Fermi constant, $C_{\mathrm{1i}}$ is a weak interaction coupling, the index $\mathrm{i=p,n}$ labels protons and neutrons, and $\mathrm{\gamma_5}$ is the fifth Dirac matrix. 
The function $\delta(\mathbf{r})$ denotes that Eq.\,(\ref{eq:Hamiltonian1}) describes a contact interaction. 
The couplings to the proton and neutron are, respectively, $C_{\mathrm{1p}}=\frac{1}{2}(1-4\sin^2{\theta_{\rm W}})$, and $C_{\mathrm{1n}}=-1/2$, with  $\theta_{\rm W}$ being the Weinberg angle. 
Treating nucleons nonrelativistically, and adding all contributions for a nucleus with $N$ neutrons and $Z$ protons, one obtains an effective NSI Hamiltonian:

\begin{equation}
   \mathrm{H^{NSI} = \frac{G_F}{\sqrt{2}}\frac{\mathit{Q}_W}{2} \gamma_5 \rho(r)}\delta(\mathbf{r}) \;,
    \label{eq:Hamiltonian2}
\end{equation}

where $Q_{\rm W}=-N+Z(1-4\sin^2{\theta_{\rm W}})$
is the weak nuclear charge, expressed here without inclusion of radiative corrections, and $\rho(r)$ is the nuclear density, assumed to be the same for protons and neutrons. As the value of $\sin^2{\theta_{\rm W}}$ is close to 1/4, there is little contribution from protons to $Q_{\rm W}$. The interaction of Eq.\,(\ref{eq:Hamiltonian2}) mixes atomic levels of opposite parity, inducing an APV observable. The atomic matrix element for this mixing needs to be calculated to extract $Q_{\rm W}$ from an experiment and determine $\sin^2{\theta_{\rm W}}$ for comparison with the SM prediction.  Thus, not only high measurement accuracy, but also accurate atomic calculations are required for APV to provide a value for $\sin^2{\theta_{\rm W}}$. 
\paragraph{}
Several accelerator experiments have extracted the Weinberg angle at various energies (see Fig.\,\ref{fig:weinberg}), providing checks of the electroweak sector of the SM and constraining beyond-SM scenarios (see, for example, \cite{E158PRL2005-Moller, Androic2018PrecisionProton, WangNature2014PVDIS, ZellerPRL2024-NuTeV}, and \cite{PDG2024PRD2024} for a summary of $\sin^2\theta_W$ determinations).
To date, the most precise low-energy determination comes from APV in Cs, via a measurement of the Cs nucleus weak charge. Not only the remarkable experimental precision in the measurement  (0.35\% in the APV effect \cite{Wood1997MeasurementCesium}) was crucial to this, but also the high precision atomic calculations achieved \cite{PorsevPRL2009,dzu12}, owing to the relatively simple Cs electronic structure, possessing a single valence electron above a tight Xe-like core. While accurate measurements  have been achieved in other heavy atoms ($\approx1\%$ in Pb \cite{Meekhof1993High-precisionLead} and Tl \cite{Vetter1995PreciseThallium}, and 0.5\% in Yb \cite{Ant19a}), the lack of precise enough atomic theory precludes competitive determinations of $\sin^2{\theta_{\rm W}}$ (see, for example, \cite{Roberts2015ParitySystems} for a summary of calculations). In the case of Yb, with two ground-level valence electrons in a $6s$ configuration that lies above a  $4f$ sub-shell with 14 electrons, atomic theory is at the $\approx 10\%$ level of accuracy \cite{Porsev1995ParityYtterbium, dzu11}. While application of neural-network tools to atomic calculations\,\cite{SafronovaArxiv2024ML} may eventually reduce theory uncertainty sufficiently, isotopic comparison of the APV effect in Yb (in which many stable isotopes are available) is free of the need for atomic calculations, and if precise enough, may produce a new determination of $\sin^2{\theta_{\rm W}}$ at low energy.

\paragraph{}

\begin{figure}{}
    \centering
    \includegraphics[width=0.8\linewidth]{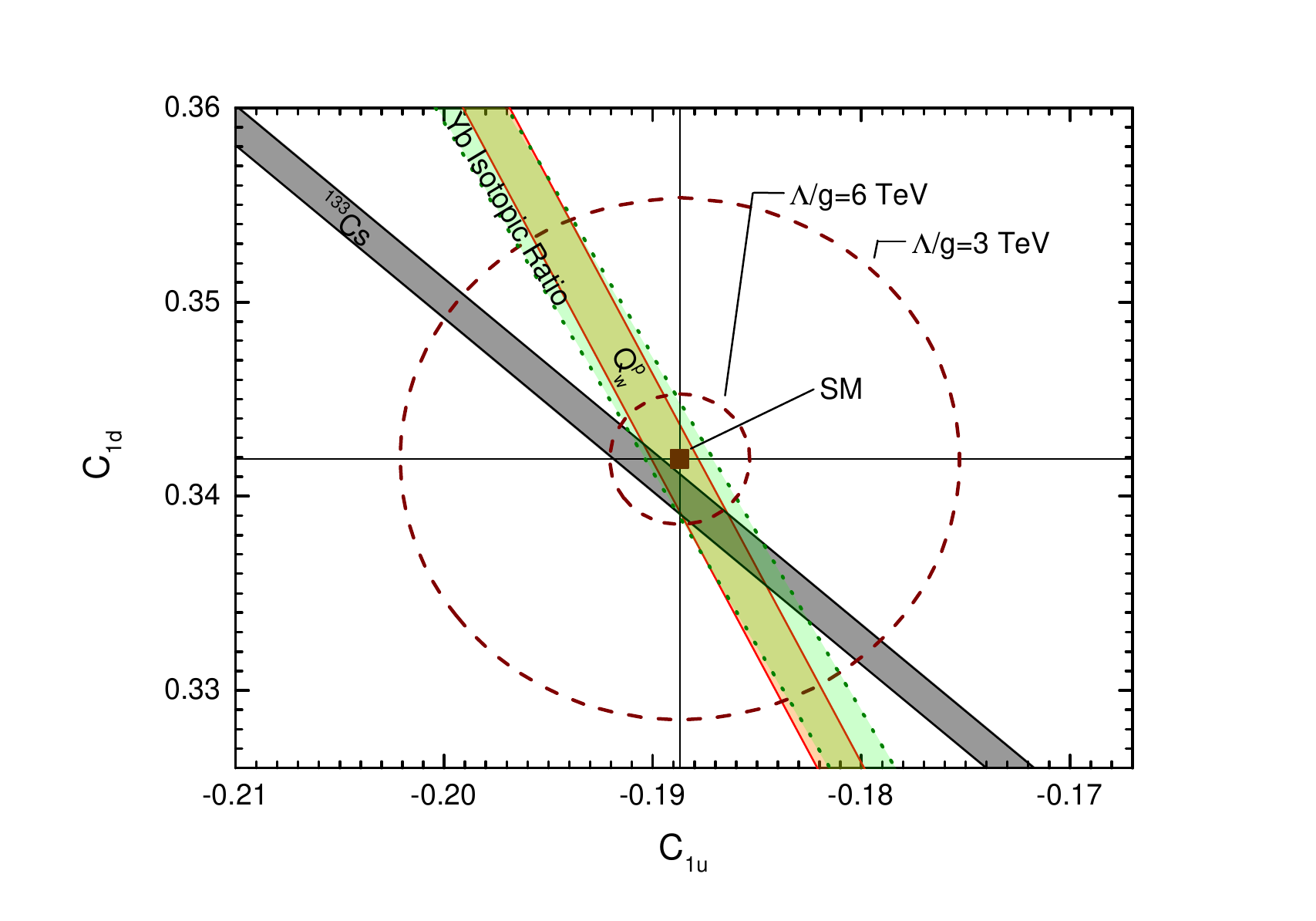}
    \caption{Constraints at the $1\sigma$ level on electron-quark couplings $C_{\textrm{1u}}$ and $C_{\textrm{1d}}$ from PV electron scattering\,\cite{Androic2018PrecisionProton} (olive), APV in Cs\,\cite{Wood1997MeasurementCesium} (gray), along with a projection from forthcoming Yb APV isotopic comparison (green). The Yb band represents the SM prediction for a 0.03\% ratio measurement in isotopes with $A=170$ and 176. Dashed circles indicate "new physics" at mass scales $\Lambda / g$, centered around the SM prediction\,\cite{Patrignani_2016}.}
    \label{fig:c1q}
 \end{figure}

The Weinberg angle is not the only parameter to be determined through a $Q_W$ extraction. The weak charge in Eq.\,(\ref{eq:Hamiltonian2}) can be expressed as:

\begin{equation}
Q_{W}=2\big(Z C_{\mathrm{1p}}+NC_{\mathrm{1n}}\big) \
\end{equation}

or, taking into account the composition of protons and neutrons in terms of up and down quarks:

\begin{equation}
Q_{W}=2\big[C_{\mathrm{1u}}(2Z+N)+C_{\mathrm{1d}}(Z+2N)\big] \;,
\end{equation}

where  $C_{\mathrm{1p}}=2C_{\mathrm{1u}}+C_{\mathrm{1d}}$ and $C_{\mathrm{1n}}=C_{\mathrm{1u}}+2C_{\mathrm{1d}}$, with $C_{\mathrm{1u}}$, $C_{\mathrm{1d}}$ being the electron couplings to the up and down quarks, respectively.  Experiments probing the electron-quark weak interaction constrain different combinations of $C_{\mathrm{1u}}$, $C_{\mathrm{1d}}$, and  only when respective constraints are combined together, is a determination of these couplings possible. Thus, APV measurements  can provide important data and complement experiments in accelerators. 
Fig.\,\ref{fig:c1q} shows how $C_{\mathrm{1u}}$ and $C_{\mathrm{1d}}$ may be determined by combining data from Cs APV \cite{Wood1997MeasurementCesium} and the Qweak experiment\,\cite{AndroicNature2018-QWEAK}.

\subsection{Nuclear-spin dependent APV}

In atoms with nuclear spin, there are additional, nuclear-spin-dependent contributions to the APV effect. The effective NSD interaction can be expressed as:

\begin{equation}
   \mathrm{H^{NSD}} =\frac{G_F}{\sqrt{2}}\kappa\frac{\boldsymbol{\alpha}\cdot\boldsymbol{\mathrm{I}}}{\mathrm{I}}\rho(r) \;,
    \label{eq:HamiltonianNSD}
\end{equation}

 where $\mathrm{\boldsymbol{\alpha}}$ is the velocity operator for atomic electrons. and $\boldsymbol{\mathrm{I}}$ is the nuclear spin. There are  three primary effects entering the NSD Hamiltonian (and the coupling $\kappa$) \cite{Roberts2015ParitySystems}: i) a $Z^0$ exchange due to the nucleon axial-vector current (Fig.\ref{fig:feynman}c).   
Only a valence (unpaired) nucleon contributes to this process (unlike the NSI $Z^0$-exchange where all nucleons contribute coherently), and does so with a small coupling, so that in heavy atoms the contribution is $\approx~10^3$ times smaller than the NSI effect. ii) 
A $Z^0$ process perturbed by the hyperfine interaction (Fig.\ref{fig:feynman}d), that is also suppressed by several orders
relative to the NSI $Z^0$-exchange. 
iii) The nuclear anapole moment (Fig.\ref{fig:feynman}e). It is a P-odd, EM moment arising due to intranuclear (hadronic) weak interactions, and in heavy atoms it dominates the overall NSD effect. 
\paragraph{}
The anapole moment was considered by Zel'dovich\,\cite{ZelDovichY1957ElectromagneticViolation}, with Flambaum \textit{et al.}\,\cite{Flambaum1980V.V.1980, FlambaumPhysLettB1984} later showing that its size grows with the atom mass number, so that it could be observed in heavy atoms. An anapole extraction can be done by checking the dependence of the APV signal on the atomic spin (i.e. by comparing APV amplitudes for different hyperfine levels of a transition).  
\paragraph{}

An anapole measurement may be used to get insight into weak interactions between hadrons and the mechanism underlying intranuclear weak forces \cite{hax13}.
The related weak-interaction processes have been formulated within the model of Desplanques, Donoghue, and Holstein (DDH model) \cite{des80} in a single-meson-exchange picture. The possible interactions are characterized through a set of weak meson–nucleon coupling constants. 
 Since the formulation of the DDH model in 1980, there have been just a handful of measurements constraining the DDH couplings, including the Cs APV experiment \cite{Wood1997MeasurementCesium}, which yielded the Cs nucleus anapole. To date, the DDH constants and, therefore, 
 the mechanisms underlying the intranuclear weak forces are not well understood. For example, there have been efforts to determine the DDH coupling $h^1_{\pi}$ (see discussion in \cite{hax13}), which relates to the pion-mediated nucleon–nucleon interaction. Current experimental data and calculations point to a suppression of the coupling, relative to the ``best guess" value provided originally by DDH \cite{des80}. Specifically, the NPDGamma collaboration \cite{bly18} measured the $\gamma$-ray asymmetry in polarized neutron capture by protons, to isolate for the first time $h^1_{\pi}$, which it determined to be $h^1_{\pi}
=(2.6\pm 1.2)\cdot10^{-7}$. In addition, most recent lattice QCD calculations yielded a   value of $h^1_{\pi}=(1.1\pm 0.5)\cdot 10^{-7}$\,\cite{hax13,was12}. Both these determinations of the coupling are in relative agreement and smaller than the DDH value of $4.6\cdot 10^{-7}$\,\cite{des80}. Still, further investigations are needed to precisely determine the strength of the pion-meadiated nucleon-nucleon weak force. 

\paragraph{}
Yb is a good candidate for  anapole moment measurements, as it possesses two stable isotopes with nuclear spin ($A=171$ with $I=1/2$, $A=173$ with $I=5/2$). As the magnitude of the Yb anapole moment is sensitive to the same coupling constant $h^1_{\pi} $ \cite{FadeevPRC2019}, its measurement is expected to offer an independent determination of the particular coupling, and guide further theoretical work on the nucleon–nucleon weak interactions. Yb atomic structure calculations \cite{sin99,por00,dzu11} have provided estimates of the nuclear-spin-dependent contributions to the APV effect. When combined with the best current values for the DDH model couplings \cite{des80}, spin-dependent contributions are expected to be around 0.1\% of the total APV effect. Therefore, achieving a measurement precision of 0.05\% in the APV effect should be sufficient to observe anapole moments.

\subsection{APV and isotopic comparison}

The experimentally determined parameter in NSI APV is the transition amplitude $E_1^{APV}=k_{PV}\cdot Q_W$, where $k_{\rm PV}$ is a factor related to the weak-interaction-induced mixing of opposite-parity atomic levels, that is almost independent of nuclear parameters. Extracting $Q_W$ requires an accurate calculation of $k_{\rm PV}$. While high precision calculations have been achieved for Cs ($\approx\,0.5\%$\cite{PorsevPRL2009, dzu12}), they have not reached comparable  precision in other heavy atoms employed thus far in APV, particularly in atoms with more than one valence electron. Considering the challenges in these calculations, it was proposed in  \cite{Dzuba1986EnhancementAtoms} to make measurements in two isotopes of the same element, and form the ratio of amplitudes so that the factor $k_{PV}$ would cancel out, yielding the ratio of the respective weak charges:

\begin{equation}
   R=\frac{Q'_W}{Q_W} \; .
    \label{eq:IsotopeRatio}
\end{equation}

With a high enough measurement accuracy, the ratio could be used to probe new physics. The sensitivity of this method to new physics was studied in \cite{MurrayWusolfPRC1999, Bro09, Via19}. While single-isotope APV is primarily sensitive to a beyond-SM electron-neutron coupling, isotope ratios are mostly sensitive to an electron-proton coupling \cite{MurrayWusolfPRC1999}. Thus, the method provides a complementary way to check for new physics. It  may  also be useful for a determination of $\sin^2{\theta_W}$. For example, measurements in a chain of four Yb isotopes $(A=170,172,174,176)$ \cite{Ant19a, Ant19b}  with single-isotope accuracy at the $0.5\%$ level, yielded several isotopic ratios that may be interpreted as a value of  $\sin^2{\theta_W}\approx 0.258\pm0.052$ \cite{PDG2022}. Forthcoming high-precision ratio determinations in Yb should reduce uncertainty, provided that systematic effects in different isotope measurements cancel out in the ratio $R$.  
\paragraph{}
Fortson \textit{et al.}  pointed out a limitation to the ratio method \cite{Fortson1990Nuclear-structureNonconservation}: $R$ is not entirely free of nuclear structure uncertainties, primarily of these in neutron distributions. As the electron-nucleus $Z^0$-exchange depends on the overlap of the electronic wavefunction $f(r)$ with the neutron and proton distributions, [$\rho_n(r)$ and $\rho_p(r)$, respectively] the weak charge is probed in a form: 

\begin{equation}
   \tilde{Q}_W=-Nq_n+Zq_p(1-4\sin^2{\theta_W}) \;,
    \label{eq:Qweaktilda}
\end{equation}

with the overlap integrals:

\begin{equation}
  q_{n(p)}=\int \rho_{n(p)}(r)f(r)d^3r \; .
    \label{eq:qfactors}
\end{equation}

In heavy atoms with neutron-rich nuclei, there is a density difference $\delta\rho=\rho_n(r)-\rho_p(r)$, that varies from one isotope to another. While proton distributions are well understood (from scattering experiments and isotope-shift spectroscopy \cite{ANGELI201369}), neutron distributions are not. It is the so-called neutron skin, the difference $\Delta R_{np}$ between the root-mean-square neutron ($R_n$) and proton ($R_p$) radii, whose (poorly known) variation between isotopes introduces uncertainty in $R$.  As argued in \cite{Fortson1990Nuclear-structureNonconservation}, the sensitivity to the neutron skin hinders the ability to probe new physics, but may be used instead to check neutron distributions. This may complement other efforts measuring neutron skins (see, for example, recent results of parity-violation electron-scattering experiments on the neutron skin of $^{208}$Pb \cite{AdikariPRL2021-PREXII} and $^{48}$Ca\,\cite{AdikariPRL2022CREX}). This is of interest not only to nuclear physics, but also to astrophysics, as the neutron skin relates to parameters of the equation of state of neutron stars \cite{ThielJPG2019,PiekarewiczPT2019}. 

\paragraph{}
The contribution of the neutron-skin effect to the ratio of Eq.\,(\ref{eq:IsotopeRatio}) for two isotopes A and A$'$ is given by \cite{Bro09}: 

\begin{equation}
\Delta R_{n.s.} \simeq \frac{3}{7} (\alpha Z)^2 \frac{1}{R_p}(\Delta R'_{np} - \Delta R_{np}) \;, 
\label{eq:DRns}
\end{equation} 

where $\alpha\approx 1/137$ is the fine structure constant. The availability of a chain of Yb isotopes may be employed to check neutron-skin variation. Using calculated neutron skins for Yb nuclei with A=170 and A$'$=176 \cite{Bro09}, one obtains a contribution to the ratio $\Delta R_{n.s.}\approx 0.14\%$. A measurement of $R$ with sub-0.1\% accuracy will enable an observation of this neutron-skin variation. Moreover, as shown in \cite{Bro09}, neutron-skin values for different isotopes are largely correlated, leading to substantial suppression of the  related  uncertainty in the ratio. This reduces the impact of the uncertainty on using the APV isotopic comparison to probe new physics.
\paragraph{}
A preliminary isotopic comparison of the Yb APV effect has been employed to check for a new  electron-proton weak coupling,  via analysis of the Yb isotopic-comparison data of \cite{Ant19a}, using the results of calculations in \cite{dzu17}. Measuring the APV isotopic variation  with neutron number allowed to isolate the proton contribution to the effect to constrain a beyond-SM interaction mediated by a Z$'$ boson with a mass lower than 1 GeV. The sensitivity in the search is greater for a Z$'$ mass lower than 10\,keV, i.e. for a Z$'$ interaction length greater than the scale at which the electron wavefunction varies rapidly around the nucleus ($\approx0.2$\,\AA). A light Z$'$ boson is  motivated in several beyond-SM scenarios, potentially explaining anomalies observed in neutrino oscillation and B-meson experiments \cite{lia17,dat19,dat18,dat17}. Availability of future, high-precision isotopic variation data in Yb will be used for a more sensitive search for a light Z$'$.

\section{Ongoing APV experiments}
In spite of the remarkable success of the Cs APV experiment (and associated atomic theory \cite {PorsevPRL2009, dzu12}) in extracting $\sin^2 \theta_W$ at low energy, and providing an anapole moment, there are good reasons to pursue further experiments. APV has been, to-date, the only method to determine the Weinberg angle at low-energy. As discussed, it complements other experiments probing the electron-quark weak interaction. There are tensions between constraints on DDH model couplings coming from different experiments,  (see, for example, discussion in \cite{saf18}); new anapole measurements would thus help to resolve present inconsistencies. In addition, isotopic APV comparison may help refine nuclear models aiming to describe neutron distributions, considering that experimental input to these models comes from a limited number of methods \cite{PiekarewiczPT2019}.
\paragraph{}
Several efforts currently aim at probing APV effects. 
Work on an improved  Yb isotopic comparison and NSD APV is underway at the University of Crete. A new setup is expected to provide greatly enhanced statistical sensitivity relative to the earlier Yb work \cite{Ant19a, Ant19b}, enough to probe neutron-skin variation among isotopes and anapole moments. Atomic theory in Yb \cite{DeMille1995ParityYtterbium, Porsev1995ParityYtterbium, dzu11} is presently at the $\sim$10\% level of accuracy in the NSI part, however isotopic ratio measurements are largely free from the need for theory.
\paragraph{}
An experiment at TRIUMF employs Fr \cite{gwi22}, where the NSI APV effect is expected to be $\approx$18 times greater than in Cs. The relatively simple atomic structure of Fr allows for accurate theory, so that  a determination of the Fr weak charge may reach, or surpass the accuracy achieved in Cs \cite{gwi22}. Availability of spinful isotopes can be employed to extract an anapole moment. A group at Purdue is pursuing work in Cs using coherent control techniques \cite{AntypasPRA2013}. One focus is on the Cs anapole moment, which can be measured in the ground-state hyperfine transition of the atom \cite{DamitzPRA2024}. Ongoing work in Mainz employs Dy, in which the presence of two  opposite-parity levels that are closely spaced, favors a strong weak-interaction-induced mixing, and thus, an enhanced APV effect. An earlier experiment yielded a null result \cite{NguyenPRA1997}, but an atomic beam setup with upgraded sensitivity \cite{Leefer2014TowardsDysprosium} may  allow to measure APV in Dy. Experiments with  single trapped ions are being pursued at Groningen  (Ba$^+$ \cite{DijckPRA2018} and Ra$^+$ \cite{PortellaSpringer2014}) and Seattle (Ba$^+$ \cite{WilliamsPRA2013}).
\paragraph{}
Measurements in molecules are also underway. The Chicago/Yale team has demonstrated the sensitivity of a molecular beam setup built to probe NSD APV in BaF. The work focuses on measuring the Ba anapole moment and the vector-electron and  axial nucleon current.  The first experiment \cite{AltuntaPRL2018, Altuntas2018MeasuringErrors} was done using the isotopologue $^{138}$Ba$^{19}$F which has an APV effect well below detection sensitivity,  therefore, confirming apparatus calibration. The experiment demonstrated the best-to-date sensitivity to NSD effects.  

\section{Experiments with Yb}

\begin{figure}[htp]
    \centering
    \includegraphics[width=0.8\linewidth]{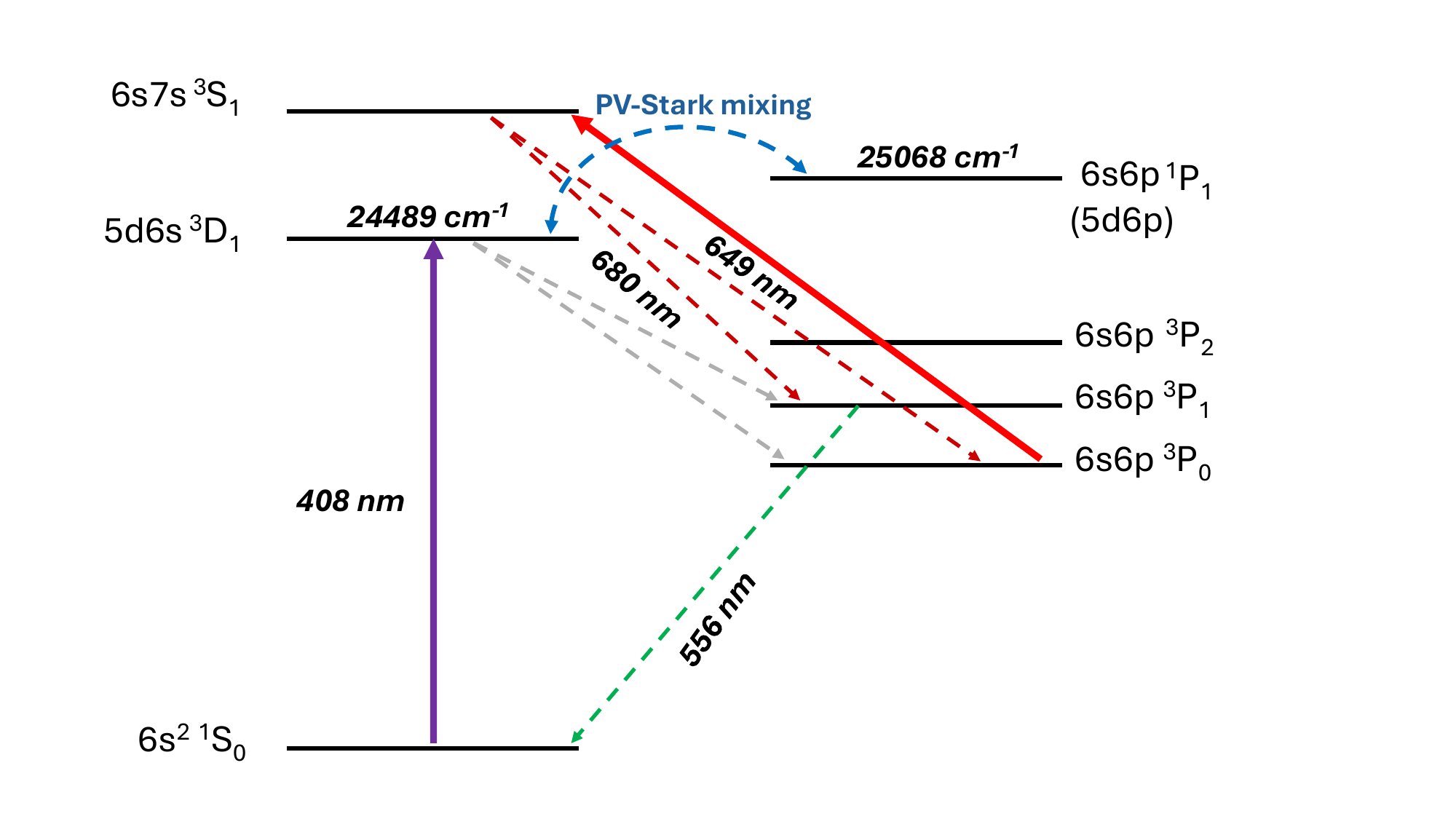}
    \caption{Partial energy level diagram of Yb showing transitions relevant to the APV experiment. Solid lines indicate excitations, while dashed lines indicate decays.}
    \label{fig:energy_diagram}
 \end{figure}

An Yb apparatus is being built at the University of Crete. This work builds on a previously developed atomic beam setup in Mainz, used to observe isotopic variation of APV \cite{Ant19a, Ant19b}. To study APV, we make use of the $\mathrm{6s^2 \; ^1S_0} \rightarrow \mathrm{5d6s \; ^3D_1}$ optical transition at 408~nm (Fig.~\ref{fig:energy_diagram}) that is E1- and M1-forbidden. Weak-force-induced mixing between the $5d6s \; ^3\mathrm{D}_1$ and $6s6p \; ^1\mathrm{P}_1$ excited levels introduces a P-odd admixture into the $^3\mathrm{D}_1$ level, and thus, a (small) E1 amplitude for the ${^1S_0} \rightarrow \mathrm{^3D_1}$ transition. Application of a slow-varying electric field $\vec{\mathrm{E}}$   introduces another (Stark) E1 amplitude, that is linear in the field. The optical field $\vec{\mathcal{E}}$ exciting atoms is linearly polarized and propagates  perpendicular to a static magnetic field $\vec{\mathrm{B}}$. The configuration of fields  $\vec{\mathrm{E}}$, $\vec{\mathcal{E}}$, $\vec{\mathrm{B}}$ is chosen to induce interference between the Stark and APV amplitudes (see Fig.\,\ref{fig:energy_diagram}). The symmetry properties of the interference term are given by the P-odd, T-even rotational invariant\,\cite{Bouchiat1986OpticalInteractions}:

\begin{equation}
    (\vec{\mathcal{E}} \cdot \vec{\mathrm{B}}) ([\vec{\mathrm{E}} \times \vec{\mathcal{E}}] \cdot \vec{\mathrm{B}}) \;.    
\end{equation}

In the absence of field imperfections, the three fields  $\vec{\mathrm{E}}$, $\vec{\mathcal{E}}$, $\vec{\mathrm{B}}$ (Fig.\,\ref{fig:energy_diagram}) are, respectively, given by: 

\begin{equation}
\vec{\mathrm{E}}=
\mathrm{E_0}\cos{\omega t}\hat{x} \; ,
\end{equation}
\begin{equation}
\vec{\mathcal{E}}=\mathrm{\mathcal{E}}(\sin\theta \hat{y}+\cos\theta\hat{z}) \; ,
\end{equation}
\begin{equation}
\vec{\mathrm{B}}=\mathrm{B}\hat{z} \; .
\end{equation}

The optical field is at an angle $\theta$ with respect to the magnetic field. The field $\vec{\mathrm{E}} $ oscillates at frequency $\omega/2\pi\approx 20$\,Hz, providing the primary reversal used to extract the Stark-PV interference from the transition rate. This rate is mainly measured  between the $m=0\rightarrow 0$ Zeeman sublevels of the $^1S_0$ and $^3D_1$ levels, and is given by:  
 
 \begin{equation}    R_{0}\propto|A_{0}^{Stark}+A_{0}^{PV}|^2=R_{0}^{[0]}+R_{0}^{[1]}\cos\omega t+R_{0}^{[2]}\cos2\omega t,
\end{equation}

where the amplitudes $R_0^{[i]}$ correspond to the different harmonics present in the rate. The $2^{\mathrm{nd}}$ harmonic with amplitude $R_0^{[2]}=    2\mathcal{E}^2\beta^2\mathrm{E_{0}}^2\sin^2\theta$ is due to the Stark effect only. The parameter $\beta$ is the transition vector polarizability \cite{Bowers1999ExperimentalYtterbium, Stalnaker2006DynamicShapes}. The $1^{\mathrm{st}}$ harmonic arises due to the Stark-PV interference and has amplitude  $R_{0}^{[1]}=8\mathcal{E}^2\beta \mathrm{E_{0}}\zeta\cos\theta\sin\theta$, where $\zeta$ is the E1 transition moment due to the weak $^3D_1-\,^{1}P_{1}$ level mixing, that contains the atomic effects of the weak force. Phase-sensitive detection at $\omega$ and $2\omega$ is done to separately extract the respective harmonic amplitudes, whose ratio is an APV-induced asymmetry:   $R_{0}^{[1]}/{R_{0}^{[2]}=\big(4\zeta/\beta \mathrm{E_0}}\big)\cot{\theta}$. This asymmetry is typically of order $ 10^{-4}$ and is further checked with a second reversal: a $\pi/2$ rotation of the angle $\theta$ between the typically used values $\pm \pi/4$. An additional (pseudo-)\,reversal is done by flipping the direction of $\vec{\mathrm{B}}$. Although this does not change the apparatus  handedness, it does provide information about systematic effects (see \cite{Ant19b}). The harmonics ratio measured under all directions of applied fields yields the ratio  $\zeta/\beta$. In \cite{Ant19a}, this was measured to be $\approx-$24\,mV/cm for $^{174}$Yb, the largest APV effect ever observed.  

\begin{figure}{}
    \centering
    \includegraphics[width=0.8\linewidth]{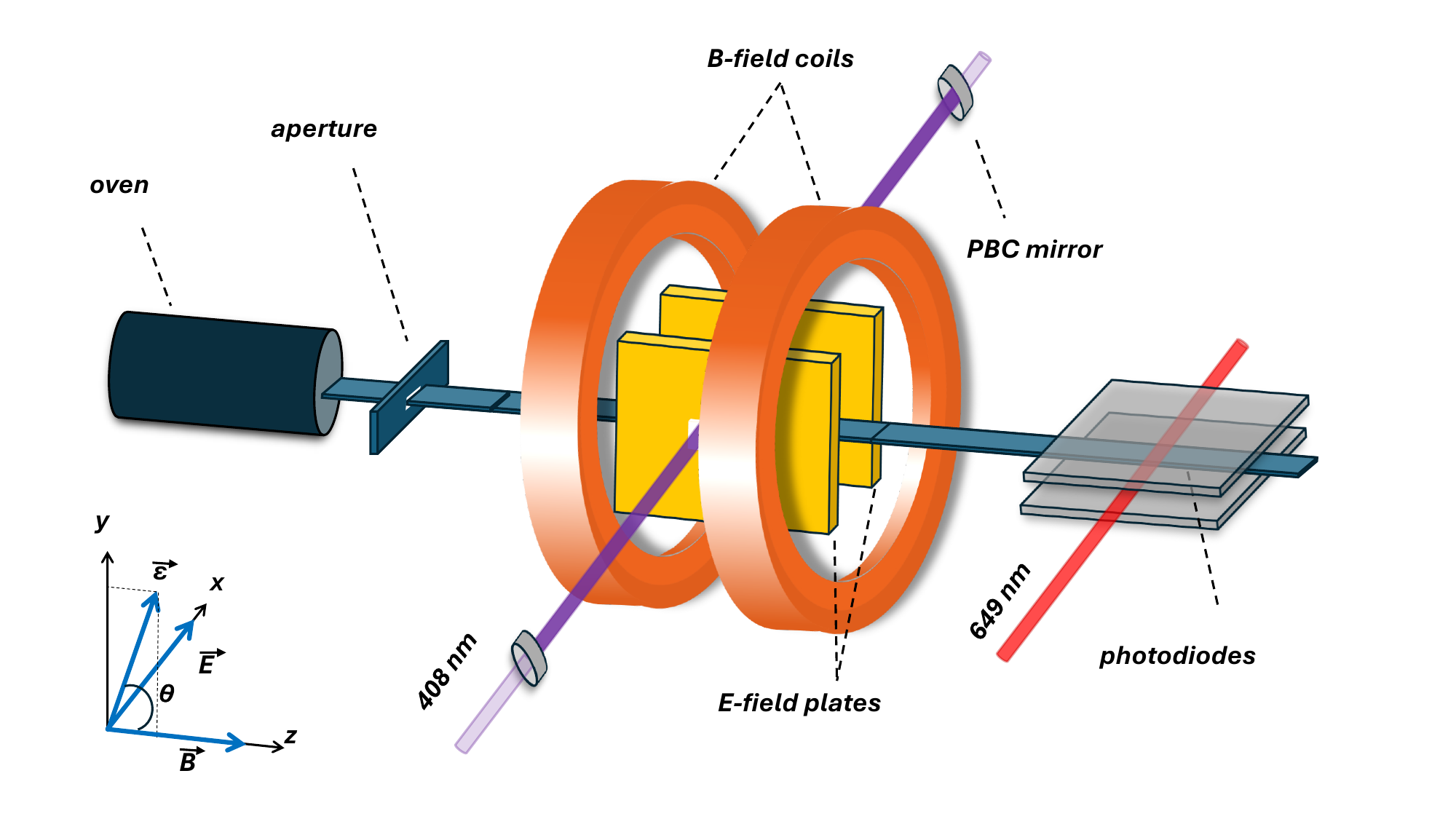}
    \caption{Schematic of the in-vacuum atomic beam apparatus.}
    \label{fig:setup}
 \end{figure}

\paragraph{}
A schematic of the atomic beam apparatus is shown in Fig.~\ref{fig:setup}. A high-flux, in-vacuum beam is formed by atoms effusing from a stainless-steel oven, that is heated to a temperature of up to $650\,\mathrm{^{\circ}C}$. After passing through several slits, atoms enter an interaction region, where  they are excited with a 408~nm optical field, circulating in a power buildup cavity (PBC) which provides about 100\,W of circulating power. The oscillating field $\vec{\mathrm{E}}$ of typical amplitude $\sim 1$\,kV/cm is produced primarily with two gold-coated, parallel field-plates. A set of Helmholtz coils is used to apply the magnetic field $\vec{\mathrm{B}}$, with $\vec{\mathrm{\rm |B|}}\approx$\,100 G. A number of auxiliary electrodes surrounding the main field plates and two sets of coils (not shown in Fig.\,\ref{fig:setup}) are used to produce, respectively, additional dc or oscillating electric fields, and auxiliary magnetic fields, for checks and control of systematics.   

\paragraph{}
Approximately 2/3 of atoms excited to the $^3\mathrm{D}_1$ level decay to the metastable $^3\mathrm{P}_0$ level from which, downstream from the interaction region, are further excited to the $\mathrm{6s7s} \; ^3\mathrm{S}_1$ state, with light at 649 nm. Fluorescence at 649 nm, 680 nm and 556 nm produced via subsequent decays (Fig.\,\ref{fig:energy_diagram}) is collected and measured, as a probe of the excitation rate in the interaction region.  In \cite{Ant19a}, fluorescence collection in this detection region was done with a set of parabolic reflectors. In the present setup these have been replaced with a pair of large-area photodiodes (of active area $48\times 48$\,mm$^2$) that are placed within 1 cm of the fluorescence column. Signal collection is much more efficient, with a $\approx15\times$ greater photocurrent. 
This enhancement, combined with the ability to produce $\approx2\times$ higher flux with the Yb oven, relative to \cite{Ant19a}, offers a substantial sensitivity enhancement. For example, the $\approx\,$440 h of the isotopic comparison of \cite{Ant19a} should be possible to carry out within less than a day. 
\paragraph{}
In \cite{Ant19a}, a total of $\approx240$\,h of data taking was required to measure both isotopes A=172 and 176 with $\approx0.4\%$ statistical accuracy. Since the detection of APV is limited by the shot-noise of atomic excitations \cite{Ant19b}, one expects that in the forthcoming experiments $\approx 1500$\,h will be sufficient to measure the effect ratio to within 0.04\%. This would allow for a $\approx2\sigma$ measurement of the neutron-skin variation between the two isotopes, that is estimated to contribute 0.09\% to the ratio\,\cite{Bro09}.

\paragraph{}

Sufficiently good experimental control of systematic effects is necessary. Such effects are generally i) signal contributions that mimic the APV signal, or ii) apparatus calibrations. APV-mimicking effects generally result from combinations of field imperfections, such as field misalignments, non-reversing field components, etc. Analysis of these contributions, and ways to measure and control them were studied in detail in \cite{Ant19b} for the case of spin-zero Yb isotopes. With moderate effort, it has been  possible to suppress these contributions to $\approx0.03\%$ of the APV effect. Moreover, they remain rather stable during a long data acquisition routine. Apparatus calibrations require a little more effort to control. The main source of uncertainty is due to the imperfect ability to measure the polarization state of the 408-nm field during an acquisition run. In the data of \cite{Ant19a} this contributed with a $0.1\%$ systematic uncertainty. An upgraded polarimeter is being employed to reduce this systematic. Other calibrations, such as, for example, uncertainty in the voltages applied to the electric field plates are below $0.1\%$.
\paragraph{}
An advantage of the isotopic ratio method is the reduced sensitivity to systematics. For the spin-zero Yb isotopes these are expected to be mostly common in different isotope measurements and thus cancel out in the ratio. While there are known systematics dependent on the isotope mass (and atomic velocity), we expect that their difference between isotopes will be negligible. 
The improved sensitivity of the newly built apparatus will allow to check this thoroughly.  Possibilities for systematics in future NSD APV studies, where the effect is compared in different hyperfine components of the 408-nm transition (for example, in the $^{171}$Yb $F=1/2\rightarrow 1/2$\, and  $F=1/2\rightarrow 3/2$ transitions, where $F$ denotes the hyperfine level) require careful consideration. It is conceivable that hyperfine-transition-dependent effects may be appreciable. A good amount of related studies were carried out in the Cs experiment, done in an atomic beam using the Stark-interference method, like the Yb experiment. Identified hyperfine-level-dependent systematics were always related to the presence of an imperfectly suppressed $\mathrm{ E^{APV}_1\cdot M1}$ interference. As the M1 transition is far better suppressed in the Yb apparatus (see discussion in\,\cite{Ant19b}), we do not expect that the NSD systematics encountered in the Cs experiment will impact our experiments. Nevertheless, a thorough study will be carried out, and it will require more focused effort relative to the NSI measurements. 

\section{Conclusions and outlook}
 
 In this article we provided a brief overview of the field of APV. We  presented the mechanisms through which the weak force induces atomic observables, and discussed the implications to nuclear and particle physics. In addition, we discussed ongoing experimental efforts including our forthcoming APV measurements in Yb.
\paragraph{}
Undoubtedly, APV studies are useful probes of the electroweak sector of the SM, providing, for example, determinations of the Weinberg angle at low energy, or complementing accelerator experiments in determining weak electron-quark couplings. For NSI measurements in a single isotope,  the requirement for precise atomic theory in order to extract electroweak parameters, limits the candidate elements to just a handful. Aside from the Cs experiment (where theory accuracy \cite{dzu12} is in pace with experiment\,\cite{Wood1997MeasurementCesium}), forthcoming measurements in Fr (where theory is almost as accurate as in Cs) appear promising \cite{gwi22} for a weak charge extraction. For the case of Yb, where sub-1\%  measurements are available\,\cite{Ant19a},  current theory accuracy severely limits interpretation in terms of SM parameters. It is possible that in forthcoming years, use of neural-network tools \cite{SafronovaArxiv2024ML} will mitigate challenges in calculating complex atoms and give a boost to the field of APV.
\paragraph{}
Lack of accurate theory in Yb may (to a certain extent) be circumvented via APV isotopic comparison. An order of magnitude better measurement accuracy is needed to extract $\sin^2\theta_{W}$ through an isotopic ratio relative to the single-isotope case, however, due to the largeness of the Yb effect and the improved sensitivity of the current apparatus, such high precision appears to be within reach. The neutron-skin related uncertainty in extracting $\sin^2\theta_{W}$ is expected to be a limiting factor at the $\sim$1\%-level in $\sin^2\theta_{W}$. As nuclear models are refined, aided by newly available neutron-skin data (\cite{AdikariPRL2021-PREXII, AdikariPRL2022CREX})  this uncertainty may be further reduced. Alternatively, the isotopic ratio may be employed to probe the variation of neutron distributions in Yb, and add to the available set of data as input to nuclear modelling. To this end, data from the forthcoming Fr measurements may also be used.  
\paragraph{}
Hadronic weak interactions can be more comfortably studied in a variety of systems, as  requirements for theory accuracy in this case are less severe. A number of ongoing experimental efforts, in Yb (in our laboratory), Cs\,\cite{DamitzPRA2024}, Fr\,\cite{gwi22} and BaF\,\cite{AltuntaPRL2018}, aim to provide new anapole moment measurements. Crucially, these studies involve isotopes possessing either valence protons or neutrons, such that the different measurements will probe different combinations of weak-meson nucleon coupling constants.
\paragraph{}

\medskip
\textbf{Acknowledgements} \par %delete if not applicable))
We acknowledge support by the European Research Council (ERC) under the European Union Horizon 2020 research and innovation program (grant agreement No 947696).

% References
\medskip

\clearpage

\bibliographystyle{MSP}
\bibliography{APVbib}

\begin{thebibliography}{10}
\providecommand{\url}[1]{\texttt{#1}}
\providecommand{\urlprefix}{URL }

\bibitem{Shi12}
V.~D. Shiltsev,
\newblock \emph{Phys.-Usp.} \textbf{2012}, \emph{55}, 10 965.

\bibitem{Gra21}
H.~M. Gray,
\newblock \emph{Rev. Phys.} \textbf{2021}, \emph{6} 100053.

\bibitem{ATL12}
{ATLAS Collaboration},
\newblock \emph{Phys. Lett. B} \textbf{2012}, \emph{716}, 1 1.

\bibitem{CMS12}
{CMS Collaboration},
\newblock \emph{Phys. Lett. B} \textbf{2012}, \emph{716}, 1 30.

\bibitem{Ber75}
J.~Bernstein,
\newblock \emph{Rev. Mod. Phys.} \textbf{1974}, \emph{46} 7.

\bibitem{ArbeyPPNP2021}
A.~Arbey, F.~Mahmoudi,
\newblock \emph{Prog. Part. Nucl. Phys.} \textbf{2021}, \emph{119} 103865.

\bibitem{saf18}
M.~Safronova, D.~Budker, D.~DeMille, D.~F.~J. Kimball, A.~Derevianko, C.~W. Clark,
\newblock \emph{Rev. Mod. Phys.} \textbf{2018}, \emph{90}, 2 025008.

\bibitem{ACM18}
{ACMe collaboration},
\newblock \emph{Nature} \textbf{2018}, \emph{562} 355.

\bibitem{Rou22}
T.~S. Roussy, L.~Caldwell, T.~Wright, W.~B. Cairncross, Y.~Shagam, K.~B. Ng, N.~Schlossberger, S.~Y. Park, A.~Wang, J.~Ye, E.~A. Cornell,
\newblock \emph{Science} \textbf{2023}, \emph{381} 6653.

\bibitem{dem03}
M.~Dine, A.~Kusenko,
\newblock \emph{Rev. Mod. Phys.} \textbf{2003}, \emph{76} 1.

\bibitem{WuPR1957}
C.~S. Wu, E.~Ambler, R.~W. Hayward, D.~D. Hoppes, R.~P. Hudson,
\newblock \emph{Phys. Rev.} \textbf{1957}, \emph{105} 1413.

\bibitem{ZelDovich1959PARITYEFFECTS}
Y.~B. Zel'~Dovich,
\newblock \emph{J. Exp. Theor. Phys. (U.S.S.R.)} \textbf{1959}, \emph{36} 964.

\bibitem{Bouchiat1974WeakPhysics}
M.~A. Bouchiat, C.~Bouchiat,
\newblock \emph{Phys. Lett. B} \textbf{1974}, \emph{48} 111.

\bibitem{BouchiatJPF1974}
M.~A. Bouchiat, C.~Bouchiat,
\newblock \emph{J. Phys. France} \textbf{1974}, \emph{35}, 12 899.

\bibitem{Bouchiat1975ParityII}
M.~A. Bouchiat, C.~Bouchiat,
\newblock \emph{J. Phys. France} \textbf{1975}, \emph{36} 493.

\bibitem{Barkov1978ObservationTransitions}
L.~M. Barkov, M.~S. Zolotorev,
\newblock \emph{JETP Letters} \textbf{1978}, \emph{27} 357.

\bibitem{MacPherson1991PreciseBismuth}
M.~J.~D. MacPherson, K.~P. Zetie, R.~B. Warrington, D.~N. Stacey, J.~P. Hoare,
\newblock \emph{Phys. Rev. Lett.} \textbf{1991}, \emph{67} 2784.

\bibitem{Meekhof1993High-precisionLead}
D.~M. Meekhof, P.~Vetter, P.~K. Majumder, S.~K. Lamoreaux, E.~N. Fortson,
\newblock \emph{Phys. Rev. Lett.} \textbf{1993}, \emph{71} 3442.

\bibitem{Phipp1996ALead}
S.~J. Phipp, N.~H. Edwards, E.~G. Baird, S.~Nakayama,
\newblock \emph{J. Phys. B} \textbf{1996}, \emph{29} 1861.

\bibitem{Vetter1995PreciseThallium}
P.~A. Vetter, D.~M. Meekhof, P.~K. Majumder, S.~K. Lamoreaux, E.~N. Fortson,
\newblock \emph{Phys. Rev. Lett.} \textbf{1995}, \emph{74} 2658.

\bibitem{Wood1997MeasurementCesium}
C.~S. Wood, S.~C. Bennett, D.~Cho, B.~P. Masterson, J.~L. Roberts, C.~E. Tanner, C.~E. Wieman,
\newblock \emph{Science} \textbf{1997}, \emph{275} 1759.

\bibitem{Guena2005MeasurementDetection}
J.~Gu{\'{e}}na, M.~Lintz, M.~A. Bouchiat,
\newblock \emph{Phys. Rev. A} \textbf{2005}, \emph{71} 042108.

\bibitem{Ant19a}
D.~Antypas, A.~Fabricant, J.~E. Stalnaker, K.~Tsigutkin, V.~V. Flambaum, D.~Budker,
\newblock \emph{Nat. Phys.} \textbf{2019}, \emph{15} 120.

\bibitem{dzu12}
V.~Dzuba, J.~Berengut, V.~Flambaum, B.~Roberts,
\newblock \emph{Phys. Rev. Lett.} \textbf{2012}, \emph{109} 203003.

\bibitem{Haxton2001AtomicMoments}
W.~C. Haxton, C.~E. Wieman,
\newblock \emph{An. Rev. Nucl. Par. Sci.} \textbf{2001}, \emph{51} 261.

\bibitem{Dzuba1986EnhancementAtoms}
V.~A. Dzuba, V.~V. Flambaum, I.~B. Khriplovich,
\newblock \emph{Zeit. Phys. D} \textbf{1986}, \emph{1} 243.

\bibitem{Fortson1990Nuclear-structureNonconservation}
E.~N. Fortson, Y.~Pang, L.~Wilets,
\newblock \emph{Phys. Rev. Lett.} \textbf{1990}, \emph{65} 2857.

\bibitem{Bro09}
B.~A. Brown, A.~Derevianko, V.~V. Flambaum,
\newblock \emph{Phys. Rev. C} \textbf{2009}, \emph{79} 035501.

\bibitem{Bouchiat1986OpticalInteractions}
M.~Bouchiat, L.~Pottier,
\newblock \emph{Science} \textbf{1986}, \emph{234} 1203.

\bibitem{LintzEL2007}
M.~Lintz, J.~Gu{\'e}na, M.-A. Bouchiat,
\newblock \emph{Eur. Phys. J. A} \textbf{2007}, \emph{32} 4.

\bibitem{Conti1979PreliminaryThallium}
R.~Conti, P.~Bucksbaum, S.~Chu, E.~Commins, L.~Hunter,
\newblock \emph{Phys. Rev. Lett.} \textbf{1979}, \emph{42} 343.

\bibitem{Tsigutkin2009ObservationYtterbium}
K.~Tsigutkin, D.~Dounas-Frazer, A.~Family, J.~E. Stalnaker, V.~V. Yashchuk, D.~Budker,
\newblock \emph{Phys. Rev. Lett.} \textbf{2009}, \emph{103} 071601.

\bibitem{WarringtonEL1993}
R.~B. Warrington, C.~D. Thompson, D.~N. Stacey,
\newblock \emph{Eur. Lett.} \textbf{1993}, \emph{24} 8.

\bibitem{Edwards1995PreciseThallium}
N.~H. Edwards, S.~J. Phipp, P.~E. Baird, S.~Nakayama,
\newblock \emph{Phys. Rev. Lett.} \textbf{1995}, \emph{74} 2654.

\bibitem{Ell17}
J.~P. Ellis,
\newblock \emph{Comp. Phys. Comm.} \textbf{2017}, \emph{210} 103.

\bibitem{Roberts2015ParitySystems}
B.~Roberts, V.~Dzuba, V.~Flambaum,
\newblock \emph{Ann. Rev. Nucl. Part. Sci.} \textbf{2015}, \emph{65} 63.

\bibitem{KhriplovichPNCbook}
I.~B. Khriplovich,
\newblock \emph{{Parity Nonconservation in Atomic Phenomena}},
\newblock London: Gordon and Breach, \textbf{2015}.

\bibitem{E158PRL2005-Moller}
{SLAC E158 Collaboration},
\newblock \emph{Phys. Rev. Lett.} \textbf{2005}, \emph{95} 081601.

\bibitem{Androic2018PrecisionProton}
D.~Androi{\'{c}}, et~al.,
\newblock \emph{Nature} \textbf{2018}, \emph{557} 207.

\bibitem{WangNature2014PVDIS}
{The Jefferson Lab Qweak Collaboration},
\newblock \emph{Nature} \textbf{2014}, \emph{506} 7486.

\bibitem{ZellerPRL2024-NuTeV}
{NuTeV Collaboration},
\newblock \emph{Phys. Rev. Lett.} \textbf{2002}, \emph{88} 091802.

\bibitem{PDG2024PRD2024}
{Particle Data Group Collaboration},
\newblock \emph{Phys. Rev. D} \textbf{2024}, \emph{110} 030001.

\bibitem{PorsevPRL2009}
S.~G. Porsev, K.~Beloy, A.~Derevianko,
\newblock \emph{Phys. Rev. Lett.} \textbf{2009}, \emph{102} 181601.

\bibitem{Porsev1995ParityYtterbium}
S.~Porsev, Y.~G. Rakhlina, M.~Kozlov,
\newblock \emph{JETP Letters} \textbf{1995}, \emph{61} 459.

\bibitem{dzu11}
V.~Dzuba, V.~Flambaum,
\newblock \emph{Phys. Rev. A} \textbf{2011}, \emph{83} 042514.

\bibitem{SafronovaArxiv2024ML}
P.~Bilous, C.~Cheung, M.~Safronova,
\newblock {(Preprint) arXiv:2408.00477, submitted: Aug 2024}.

\bibitem{AndroicNature2018-QWEAK}
{The Jefferson Lab Qweak Collaboration},
\newblock \emph{Nature} \textbf{2018}, \emph{557}, 7704.

\bibitem{Accardi2016}
A.~Accardi, J.~L. Albacete, M.~Anselmino, N.~Armesto, E.~C. Aschenauer, A.~Bacchetta, D.~Boer, W.~K. Brooks, T.~Burton, N.-B. Chang, W.-T. Deng, A.~Deshpande, M.~Diehl, A.~Dumitru, R.~Dupr{\'e}, R.~Ent, S.~Fazio, H.~Gao, V.~Guzey, H.~Hakobyan, Y.~Hao, D.~Hasch, R.~Holt, T.~Horn, M.~Huang, A.~Hutton, C.~Hyde, J.~Jalilian-Marian, S.~Klein, B.~Kopeliovich, Y.~Kovchegov, K.~Kumar, K.~Kumeri{\v{c}}ki, M.~A.~C. Lamont, T.~Lappi, J.-H. Lee, Y.~Lee, E.~M. Levin, F.-L. Lin, V.~Litvinenko, T.~W. Ludlam, C.~Marquet, Z.-E. Meziani, R.~McKeown, A.~Metz, R.~Milner, V.~S. Morozov, A.~H. Mueller, B.~M{\"u}ller, D.~M{\"u}ller, P.~Nadel-Turonski, H.~Paukkunen, A.~Prokudin, V.~Ptitsyn, X.~Qian, J.-W. Qiu, M.~Ramsey-Musolf, T.~Roser, F.~Sabati{\'e}, R.~Sassot, G.~Schnell, P.~Schweitzer, E.~Sichtermann, M.~Stratmann, M.~Strikman, M.~Sullivan, S.~Taneja, T.~Toll, D.~Trbojevic, T.~Ullrich, R.~Venugopalan, S.~Vigdor, W.~Vogelsang, C.~Weiss, B.-W. Xiao, F.~Yuan, Y.-H. Zhang, L.~Zheng,
\newblock \emph{Eur. Phys. J. A} \textbf{2016}, \emph{52}, 9.

\bibitem{dav15}
H.~Davoudiasl, H.-S. Lee, W.~J. Marciano,
\newblock \emph{Phys. Rev. D} \textbf{2015}, \emph{92} 055005.

\bibitem{dav14}
H.~Davoudiasl, H.-S. Lee, W.~J. Marciano,
\newblock \emph{Phys. Rev. D} \textbf{2014}, \emph{89} 095006.

\bibitem{Patrignani_2016}
C.~Patrignani,
\newblock \emph{Chi. Phys. C} \textbf{2016}, \emph{40}, 10 100001.

\bibitem{ZelDovichY1957ElectromagneticViolation}
{Zel' Dovich Y},
\newblock \emph{J. Exp. Theor. Phys. (U.S.S.R.)} \textbf{1957}, \emph{33} 1531.

\bibitem{Flambaum1980V.V.1980}
V.~V. Flambaum, I.~B. Khriplovich,
\newblock \emph{Zh. Eksp. Teor. Fiz.} \textbf{1980}, \emph{79} 1656.

\bibitem{FlambaumPhysLettB1984}
V.~Flambaum, I.~Khriplovich, O.~Sushkov,
\newblock \emph{Phys. Lett. B} \textbf{1984}, \emph{146} 367.

\bibitem{hax13}
W.~C. Haxton, B.~R. Holstein,
\newblock \emph{Prog. Part. Nucl. Phys.} \textbf{2013}, \emph{71} 185.

\bibitem{des80}
B.~Desplanques, J.~F. Donoghue, B.~R. Holstein,
\newblock \emph{Ann. Phys.} \textbf{1980}, \emph{124} 449.

\bibitem{bly18}
{NPDGamma Collaboration},
\newblock \emph{Phys. Rev. Lett.} \textbf{2018}, \emph{121} 242002.

\bibitem{was12}
J.~Wasem,
\newblock \emph{Phys. Rev. C} \textbf{2012}, \emph{85} 022501.

\bibitem{FadeevPRC2019}
P.~Fadeev, V.~V. Flambaum,
\newblock \emph{Phys. Rev. C} \textbf{2019}, \emph{100} 015504.

\bibitem{sin99}
Y.~Singh, D.~Giri,
\newblock \emph{J. Phys. A} \textbf{1999}, \emph{32} L407.

\bibitem{por00}
S.~Porsev, M.~Kozlov, Y.~G. Rakhlina,
\newblock \emph{Hyperfine Inter.} \textbf{2000}, \emph{127} 395.

\bibitem{MurrayWusolfPRC1999}
M.~J. Ramsey-Musolf,
\newblock \emph{Phys. Rev. C} \textbf{1999}, \emph{60} 015501.

\bibitem{Via19}
A.~V. Viatkina, D.~Antypas, M.~G. Kozlov, D.~Budker, V.~V. Flambaum,
\newblock \emph{Phys. Rev. C} \textbf{2019}, \emph{100} 034318.

\bibitem{Ant19b}
D.~Antypas, A.~M. Fabricant, J.~E. Stalnaker, K.~Tsigutkin, V.~V. Flambaum, D.~Budker,
\newblock \emph{Phys. Rev. A} \textbf{2019}, \emph{100} 012503.

\bibitem{PDG2022}
{Particle Data Group},
\newblock \emph{Prog. Theor. Exp. Phys.} \textbf{2022}, \emph{2022} 8.

\bibitem{ANGELI201369}
I.~Angeli, K.~Marinova,
\newblock \emph{At. Data Nucl. Data Tables} \textbf{2013}, \emph{99}, 1 69.

\bibitem{AdikariPRL2021-PREXII}
{PREX Collaboration},
\newblock \emph{Phys. Rev. Lett.} \textbf{2021}, \emph{126} 172502.

\bibitem{AdikariPRL2022CREX}
{CREX Collaboration},
\newblock \emph{Phys. Rev. Lett.} \textbf{2022}, \emph{129} 042501.

\bibitem{ThielJPG2019}
M.~Thiel, C.~Sfienti, J.~Piekarewicz, C.~J. Horowitz, M.~Vanderhaeghen,
\newblock \emph{J. Phys. G} \textbf{2019}, \emph{46} 9.

\bibitem{PiekarewiczPT2019}
J.~Piekarewicz, F.~J. Fattoyev,
\newblock \emph{Phys. Today} \textbf{2019}, \emph{72} 30.

\bibitem{dzu17}
V.~Dzuba, V.~Flambaum, Y.~Stadnik,
\newblock \emph{Phys. Rev. Lett.} \textbf{2017}, \emph{119} 223201.

\bibitem{lia17}
J.~Liao, D.~Marfatia,
\newblock \emph{Phys. Lett. B} \textbf{2017}, \emph{775} 54.

\bibitem{dat19}
A.~Datta, B.~Dutta, S.~Liao, D.~Marfatia, L.~E. Strigari,
\newblock \emph{J. High Energy Phys.} \textbf{2019}, \emph{2019} 1.

\bibitem{dat18}
A.~Datta, J.~Kumar, J.~Liao, D.~Marfatia,
\newblock \emph{Phys. Rev. D} \textbf{2018}, \emph{97} 115038.

\bibitem{dat17}
A.~Datta, J.~Liao, D.~Marfatia,
\newblock \emph{Phys. Lett. B} \textbf{2017}, \emph{768} 265.

\bibitem{DeMille1995ParityYtterbium}
D.~DeMille,
\newblock \emph{Phys. Rev. Lett.} \textbf{1995}, \emph{74} 4165.

\bibitem{gwi22}
G.~Gwinner, L.~Orozco,
\newblock \emph{Quant. Sci. Tech.} \textbf{2022}, \emph{7} 024001.

\bibitem{AntypasPRA2013}
D.~Antypas, D.~S. Elliott,
\newblock \emph{Phys. Rev. A} \textbf{2013}, \emph{87} 042505.

\bibitem{DamitzPRA2024}
A.~Damitz, J.~A. Quirk, C.~E. Tanner, D.~S. Elliott,
\newblock \emph{Phys. Rev. A} \textbf{2024}, \emph{109} 032810.

\bibitem{NguyenPRA1997}
A.~T. Nguyen, D.~Budker, D.~DeMille, M.~Zolotorev,
\newblock \emph{Phys. Rev. A} \textbf{1997}, \emph{56} 3453.

\bibitem{Leefer2014TowardsDysprosium}
N.~Leefer, L.~Bougas, D.~Antypas, D.~Budker,
\newblock {{(Preprint) arXiv:1412.1245, submitted: Dec 2014}}.

\bibitem{DijckPRA2018}
E.~A. Dijck, A.~Mohanty, N.~Valappol, M.~N.~n. Portela, L.~Willmann, K.~Jungmann,
\newblock \emph{Phys. Rev. A} \textbf{2018}, \emph{97} 032508.

\bibitem{PortellaSpringer2014}
M.~Nu{\~n}ez~Portela, E.~A. Dijck, A.~Mohanty, H.~Bekker, J.~E. van~den Berg, G.~S. Giri, S.~Hoekstra, C.~J.~G. Onderwater, S.~Schlesser, R.~G.~E. Timmermans, O.~O. Versolato, L.~Willmann, H.~W. Wilschut, K.~Jungmann,
\newblock \emph{Appl. Phys. B} \textbf{2014}, \emph{114} 173.

\bibitem{WilliamsPRA2013}
S.~R. Williams, A.~Jayakumar, M.~R. Hoffman, B.~B. Blinov, E.~N. Fortson,
\newblock \emph{Phys. Rev. A} \textbf{2013}, \emph{88} 012515.

\bibitem{AltuntaPRL2018}
E.~Altunta\ifmmode~\mbox{\c{s}}\else \c{s}\fi{}, J.~Ammon, S.~B. Cahn, D.~DeMille,
\newblock \emph{Phys. Rev. Lett.} \textbf{2018}, \emph{120} 142501.

\bibitem{Altuntas2018MeasuringErrors}
E.~Altuntas, J.~Ammon, S.~B. Cahn, D.~Demille,
\newblock \emph{Phys. Rev. A} \textbf{2018}, \emph{97} 042101.

\bibitem{Bowers1999ExperimentalYtterbium}
C.~J. Bowers, D.~Budker, S.~J. Freedman, G.~Gwinner, J.~E. Stalnaker, D.~DeMille,
\newblock \emph{Phys. Rev. A} \textbf{1999}, \emph{59} 3513.

\bibitem{Stalnaker2006DynamicShapes}
J.~E. Stalnaker, D.~Budker, S.~J. Freedman, J.~S. Guzman, S.~M. Rochester, V.~V. Yashchuk,
\newblock \emph{Phys. Rev. A} \textbf{2006}, \emph{73} 043416.

\end{thebibliography}

\end{document}